\documentstyle[floats,pra,aps]{revtex}
\tightenlines 

\def\beq{\begin{equation}}
\def\eeq{\end{equation}}
\def\bea{\begin{eqnarray}}
\def\eea{\end{eqnarray}}
\def\nn{\nonumber}
\def\ba{\begin{array}}
\def\ea{\end{array}}   
\def\0{{\mbox{\boldmath $0$}}}
\def\one{1\hskip -.771mm{\rm l}}
\def\A{{\mbox{\boldmath $A$}}}
\def\B{{\mbox{\boldmath $B$}}}

\def\S{{\mbox{\boldmath $S$}}}

\def\vpi{{\mbox{\boldmath $\pi$}}}
\def\r{{\mbox{\boldmath $r$}}}

\def\zi{z_{\rm in}}  
\def\zo{z_{\rm out}}

\def\ptwo{\hat{p}_\perp^{\,2}}

\def\pitwo{\hat{\pi}_\perp^{\,2}}

\def\ddz{\frac{\partial}{\partial z}} 
\def\vsig{{\mbox {\boldmath $\sigma$ }}} 
\def\Al{{\mbox{\boldmath $\alpha$}}}

\def\half{\frac{1}{2}}

\def\al{\alpha}

\def\E{{\hat{\cal E}}}
\def\O{{\hat{\cal O}}}
\def\eps{\epsilon}
\def\g{\gamma}
\def\Vomeg{{\underline{\mbox {\boldmath $\Omega$}}}_s}

\def\vpip{{\hat{\mbox{\boldmath $\pi$}}}_\perp}

\def\hH{\hat{H}} 
\def\Vsig{{\mbox {\boldmath $\Sigma$ }}} 
\def\Nab{{\mbox{\boldmath $\nabla$}}}

\begin{document}

\title{\large\bf Quantum Mechanical Formalism of Particle Beam 
Optics\footnote{\normalsize To apear in the Proceedings of 
the 18th Advanced ICFA Beam Dynamics Workshop on
Quantum Aspects of Beam Physics (QABP) (15-20 October 2000, Capri,
ITALY) Editor: Pisin Chen (World Scientific, Singapore, 2002)
http://qabp2k.sa.infn.it/
}}
\author{Sameen Ahmed KHAN}
\address
{
Dipartimento di Fisica Galileo Galilei  
Universit\`{a} di Padova \\
Istituto Nazionale di Fisica Nucleare~(INFN) Sezione di Padova \\
Via Marzolo 8 Padova 35131 ITALY \\
E-mail: khan@pd.infn.it, http://www.pd.infn.it/$\sim$khan/ \\
E-mail: rohelakhan@yahoo.com, http://www.imsc.ernet.in/$\sim$jagan/} 

\maketitle

\medskip

\noindent
{\bf Abstract:}
A general procedure for construction of the formalism of quantum beam optics
for any particle is reviewed.  The quantum formalism of spin-$\frac{1}{2}$ 
particle beam optics is presented starting {\em ab initio} with the Dirac 
equation.  As an example of application the case of normal magnetic quadrupole 
lens is discussed.  In the classical limit the quantum formalism leads to the 
well-known Lie algebraic formalism of classical particle beam optics.

\medskip

\section{Introduction} 
Whenever the possibility of a quantum formalism of particle beam optics is 
mentioned the immediate response, invariably, in the accelerator physics 
community is to  ask what is the need to use quantum mechanics when classical 
mechanics has been so successful in the design and operation of numerous 
accelerators?  Of course, this is a natural question and, though the system is 
quantum mechanical at the fundamental level, in most situations classical  
mechanics is quite adequate~\cite{C} since {\em the de Broglie wavelength of the} 
({\em high energy}) {\em beam particle is very small compared to the typical 
apertures of the cavities in accelerators} as has been pointed out clearly  
by Chen.~\cite{Chen}  But, the recent attention to the sensitivity of tracking 
of particle trajectories to quantum granularities in the stochastic regions of 
phase space~\cite{HY} and the limits placed by quantum mechanics on the 
achievable beam spot sizes in accelerators~\cite{Hill} clearly indicates the 
need for a formalism of quantum beam optics relevant to such issues.~\cite{V}  
Besides this, with ever increasing demand for higher energies and luminosity 
and lower emittance beams, and the need for polarized beams, the interest in 
the studies on the various quantum aspects of beam physics is growing.~\cite{QABP} 
So, it is time that a quantum formalism of particle beam dynamics is developed 
in which all aspects (optical, spin, radiation, $\ldots$, etc.) are considered 
in a unified framework.  

\medskip

The grand success of the classical theories accounts for the very few quantum 
approaches to the charged-particle beam optics in the past.  Notable among
these are:

\medskip

\begin{itemize}
\item
1930 Glaser: Quantum theory of image formation in electron microscopy - 
Semiclassical theory based on the nonrelativistic Schr\"{o}dinger 
equation.~\cite{HK3}  
\item
1934 Rubinovicz; 1953 Durand; 1953 Phan-Van-Loc: Studies on electron diffraction 
based on the Dirac equation.~\cite{HK3} 
\item
1986 Ferwerda {\it et al.}: Justified the use of scalar~(Klein-Gordon) equation 
for image formation in practical electron microscopes operating even at 
relativistic energies.~\cite{HK3} 
\item
1989-90 Jagannathan {\it et al.}: The first derivation of the focusing theory of 
electron lenses using the Dirac equation.~\cite{JSSM}  1995: Quantum theory of 
aberrations to all orders using the Klein-Gordon theory and the Dirac 
Theory.~\cite{KJ3} 1996: Spin dynamics of the Dirac particle beam.~\cite{CJKP} 
\end{itemize}

\medskip

The formalism of {\em quantum theory of charged-particle beam optics} developed 
by Jagannathan {\em et al.}, based on the Klein-Gordan and Dirac equations, 
provides a recipe to work out the quantum maps for any particle optical system up 
to any desired order.~\cite{JSSM}$^-$\cite{JK3}. The classical limit (de Broglie 
wavelength $\longrightarrow 0$) of this quantum formalism reproduces the well-known 
Lie algebraic approach of Dragt {\em et al.}~\cite{Lie} for handling the classical 
beam optics.  Spin evolution, independent of orbital motion, can also be 
treated classically using the Lie algebraic approach.~\cite{YEY}  This brief 
note is to present the essential features of the quantum formalism of 
spin-$\frac{1}{2}$ particle beam optics based on the Dirac equation. 

\medskip

\section{The general formalism of quantum beam optics}
In many accelerator optical elements the electromagnetic fields are static or 
can be reasonably assumed to be static.  In such devices one can further 
ignore the times of flights which may be negligible, or of no direct relevance, 
as the emphasis is more on the profiles of the trajectories.  The idea is to 
analyze the evolution of the beam parameters of the various individual 
charged-particle beam optical elements (quadrupoles, bending magnets, $\cdots$) 
along the optic axis of the system.  Let us consider a charged-particle at the 
point $({\mbox{\boldmath $r$}}_\perp , s_{\rm in})$ where 
${\mbox{\boldmath $r$}}_\perp$ is the transverse coordinate and $s$ refers to 
the coordinate along the optic axis.  After passing through the system 
this particle arrives at the point 
$({\mbox{\boldmath $r$}}_\perp , s_{\rm out})$.  Note that 
$({\mbox{\boldmath $r$}}_\perp , s)$ constitute a curvilinear coordinate 
system, adapted to the geometry of the system.  Given the initial quantities at 
an $s_{\rm in}$, the problem is to determine the final quantities at an 
$s_{\rm out}$, and to design an optical device in such a way that the relations 
between the initial and final quantities have the desired properties.  Since we 
want to know the evolution of the beam parameters along the optic axis of the 
system the starting equation of the quantum formalism should be desirably of 
the form 
\bea
i \hbar \frac{\partial }{\partial s} \psi (\r_\perp ; s) 
= \hat{\cal H}  \psi (\r_\perp ; s), 
\label{BOE}
\eea
linear in $\partial/{\partial s}$, irrespective of the basic time-dependent 
equation (Schr\"{o}dinger, Klein-Gordon, Dirac, $\cdots$) governing the system.   
So the step\,-\,I of building the quantum formalism is to cast the basic equation  
of quantum mechanics, relevant for the system under study, in the form~(\ref{BOE}).  
Once this is done the step\,-\,II would be to obtain the relationship for the 
quantities at any point $s$ to the quantities at the point $s_{\rm in}$.  This 
in the language of the quantum formalism would require to obtain the relationship 
for an observable  
$\left\{\left\langle O \right\rangle \left( s \right) \right\}$ at the transverse 
plane at $s$ to the observable 
$\left\{ \left\langle O \right\rangle \left(s_{\rm in} \right) \right\}$
at the transverse plane at $s_{\rm in}$.  This can be achieved by 
integrating~(\ref{BOE}).  Formally, 
\bea
\psi \left(\r_\perp ; s \right) & = &
\hat{U} \left(s , s_{\rm in} \right)
\psi \left(\r_\perp ; s_{\rm in} \right)\,, 
\label{BOI}
\eea
which leads to the required transfer maps
\bea
\left\langle O \right\rangle \left(s_{\rm in} \right) 
\longrightarrow
\left\langle O \right\rangle \left(s \right)
= \left\langle \psi \left(s \right) 
\left| O \right|
\psi \left(s \right) \right\rangle 
= \left\langle \psi \left(s_{\rm in} \right) 
\left| \hat{U}^{\dagger} O  \hat{U} \right|
\psi \left( s_{\rm in} \right) \right\rangle\,.
\label{BOM}
\eea
Equation~(\ref{BOE}) is the basic equation of the quantum formalism
of charged-particle beam optics and we call it as the {\em beam optical
equation}, $\hat{\cal H}$ as the {\em beam optical Hamiltonian} and
$\psi$ as the {\em beam optical wavefunction}.  

\medskip

To summarize, we have a two-step algorithm to build a quantum formalism of 
charged-particle beam optics.  On may question the applicability of the two-step 
algorithm: Does it always work?  From experience we know that it works for the 
Schr\"{o}dinger, Klein-Gordon and Dirac equations.  The above description gives 
an oversimplified picture of the formalism than, it actually is.  There are 
several crucial points to be noted to understand the success of the two-step 
algorithm.  The first step in the algorithm to obtain the beam optical equation 
is much more than a mere mathematical transformation which eliminates `$t$' in 
preference to a variable `$s$' along the optic axis.  There has to be a clever 
set of transformations ensuring that the resultant $s$-dependent equation has
a very close physical and mathematical analogy with the original $t$-dependent 
equation of the standard quantum mechanics.  Without this guiding requirement it 
would not be possible to execute the second step of the algorithm which ensures 
that we can use all the rich machinery of the quantum mechanics to compute the 
transfer maps characterizing the optical system.  This summarizes the recipe of 
obtaining the quantum prescriptions for the optical transfer maps.  Rest is 
mostly a computational affair which is inbuilt in the powerful algebraic 
machinery of the algorithm.  As in any computation, there are some reasonable 
assumptions and some possible approximations coming from physical considerations.  
It is important to note that in the case of the Schr\"{o}dinger, Klein-Gordon 
and Dirac equations the beam optical forms obtained are exact.  Approximations 
necessarily enter only in the step\,-\,II of the algorithm, {\em i.e.,} while 
integrating the beam optical equation and computing the transfer maps for the 
quantum averages of the beam observables.   As in the classical theory, the 
approximations arise due to the fact that only the first few terms are retained 
in the infinite series expansion of the beam optical Hamiltonian.  The beam 
optical Hamiltonian is  obtained as a power series in 
$\left| {\hat{\vpi}_\perp}/{p_0} \right|$ where $p_0$ is the design (or average) 
momentum of the beam particles moving predominantly along the optic axis of the 
system and $\hat{\vpi}_\perp$ is the small transverse kinetic momentum.  The 
leading order contribution gives rise to the {\em paraxial} or the ideal 
behavior and higher order contributions give rise to the nonlinear or 
{\em aberrating} behavior.  Both the paraxial and the aberrating behaviors 
deviate from their classical nature by quantum contributions which are in 
powers of the de Broglie wavelength of the beam particle 
($\lambda_0 = {2 \pi \hbar}/{p_0}$).  The classical formalism is obtained from 
the quantum formalism by taking the limit $\lambda_0 \longrightarrow 0$.  

\medskip

\section{Formalism of the Dirac particle beam optics} 
Now we shall see how the above algorithm works for the Dirac particle.  Let us 
consider a monoenergetic beam of Dirac particles particles of mass $m$, charge 
$q$ and anomalous magnetic moment $\mu_a$, transported through a magnetic optical 
element with a straight optic axis characterized by the static potentials 
$\left(\phi (\r ), \A (\r ) \right)$.  The beam propagation is governed
by the stationary Dirac equation
\beq
\hat{\rm H}_D \left| \psi_D \right\rangle = E 
\left| \psi_D \right\rangle\,,
\label{DE}
\eeq
where $\left| \psi_D \right\rangle$ is the time-independent $4$-component Dirac 
spinor, $E$ is the total energy of the beam particle and the Hamiltonian 
$\hat{\rm H}_D$, including the Pauli term, is given by
\bea 
\hat{\rm H}_D & = & \beta m c^2 + c \Al \cdot (- i\hbar \Nab - q \A)
- \mu_a \beta \Vsig \cdot \B \,, 
\eea
where the symbols have their usual meanings.~\cite{BD}  To cast (\ref{DE}) in 
the required beam optical form (\ref{BOE}) we multiply $\hat{\rm H}_D$ (on the 
left) by $\al_z/c$ and rearrange the terms to get 
\bea 
i\hbar\frac{\partial}{\partial z}\left| \psi_D \right\rangle
& = & \hat{\cal H}_D \left| \psi_D \right\rangle\,, \nn \\
\hat{\cal H}_D & = & - p_0 \beta \chi \al _z - q A_z I + \al _z \Al 
_\perp \cdot \vpip  + (\mu_a /c) \beta \al _z \Vsig \cdot \B\,, 
\label{inter} 
\eea 
where $\chi$ is a diagonal matrix with elements $(\xi,\xi,-1/\xi,-1/\xi)$ and 
$\xi = \sqrt{(E+mc^2)/(E-mc^2)}$.  Equation~(\ref{inter}) is still not in a 
completely desirable form.  So we resort to a further transformation:  
\beq
\left| \psi_D \right\rangle 
\longrightarrow
\left| \psi ' \right\rangle 
= M \left| \psi_D \right\rangle\,, \qquad 
M  = \frac{1}{\sqrt{2}} (I + \chi \al _z)\,. 
\label{Mtransf}
\eeq
Then we obtain
\beq 
i\hbar \ddz \left| \psi ' \right\rangle = \hat{\cal H} ' \left| \psi ' 
\right\rangle\,, \quad 
\hat{\cal H} ' =  M \hat{\cal H}_D M^{-1} = - p_0 \beta + \E + \O \,,
\eeq
where the nonvanishing matrix elements of the even term $\E$ and the odd term $\O$ 
are given by 
\bea 
\E _{11} & = & - q A_z \one - 
(\mu_a /2c) \left\{ \left(\xi + \xi^{-1} \right) \vsig _\perp \cdot 
\B _\perp + \left( \xi - \xi^{-1} \right) \sigma_z B_z \right\}\,,  
\nn \\
\E _{22} & = & - q A_z \one - 
(\mu_a /2c) \left\{ \left( \xi + \xi^{-1} \right) \vsig _\perp 
\cdot \B _\perp - \left( \xi - \xi^{-1} \right) \sigma_z B_z 
\right\}\,,  \nn \\ 
\O _{12} & = & \xi\,\left[ \vsig _\perp \cdot \vpip  
- (\mu_a /2c) \left\{ i \,\left( \xi - \xi^{-1} \right) \left( B_x 
\sigma_y - B_y \sigma_x \right) \right. \right. 
\nn \\ 
  &   & \phantom{- \xi^{-1}\,[ \vsig _\perp \cdot \vpip  
        + (\mu_a /2c) \{i} \left. \left.- \left( \xi + \xi^{-1} 
\right) B_z \one \right\} \right]\,, \nn \\
\O _{21} & = & - \xi^{-1}\,\left[ \vsig _\perp \cdot \vpip  
+ (\mu_a /2c) \left\{ i \,\left( \xi - \xi^{-1} \right) \left( B_x 
\sigma_y - B_y \sigma_x \right) \right. \right. \nn \\
  &   & \phantom{- \xi^{-1}\,[ \vsig _\perp \cdot \vpip 
        + (\mu_a /2c) \{ i} \left. \left. + \left( \xi + \xi^{-1} 
\right) B_z \one \right\} \right]\,.
\eea

\medskip

The effect of the transformation (\ref{Mtransf}) is to make the lower components 
of a Dirac spinor corresponding to a quasiparaxial beam moving in the positive 
$z$-direction negligible compared to the upper components and thus effectively 
making the $4$-component spinor as a $2$-component spinor.  Now one may observe 
the close analogy: 

\smallskip 

\begin{tabular}{ll}
{\it Standard Dirac equation} ~~~~~~~  & {\it Beam optical form} \\
$m c^2 \beta + \E _D + \O _D$  & $ - p_0 \beta + \E + \O $ \\
Positive energy   & Forward propagation \\
Nonrelativistic, $ c\left|\mbox{\boldmath $\pi$}\right| \ll mc^2$ &
Paraxial beam, $\left|\mbox{\boldmath $\pi$}_\perp \right| \ll p_0$ \\
$ m c^2 $: Note $i\hbar\frac{\partial \psi}{\partial t} \approx mc^2\psi$ & 
$ - p_0$: Note $i\hbar\frac{\partial \psi}{\partial z} \approx -p_0\psi$ \\
Nonrelativistic motion       & Paraxial behavior \\
~~ + Relativistic corrections & ~~ + Aberration corrections \\
\end{tabular}

\smallskip

\noindent 
This completes the step\,-\,I of the algorithm.  To execute the step\,-\,II 
we proceed as follows. 

The above analogy suggests that, as the most systematic way to understand the Dirac 
Hamiltonian as a nonrelativistic part plus relativistic correction terms is to 
use the Foldy-Wouthuysen (FW) transformation technique,~\cite{BD} we should 
adopt a similar FW-like approach in the beam optical case to understand the 
beam optical Hamiltonian as a paraxial part plus nonparaxial correction terms.  
This leads to a procedure to obtain the paraxial behavior accompanied by 
a systematic method to compute the aberrations to {\em all} orders in powers of 
the expansion parameter $1/{p_0}$.  To leading order, the first FW-like 
transformation is 
\beq 
\left| \psi^{(1)} \right\rangle = \exp \left(-\beta \O/2p_0\right) 
\left| \psi ' \right\rangle\,. 
\label{FW1}
\eeq
Then
\bea 
i\hbar \ddz \left| \psi^{(1)} \right\rangle 
& = & \hat{\cal H}^{(1)} \left| \psi^{(1)} \right\rangle\,, \qquad 
\hat{\cal H}^{(1)} = - p_0 \beta + \E ^{(1)} + \O ^{(1)}\,, \nn \\
\E ^{(1)} = \E - \frac{1}{2 p_0} \beta \O ^2 & + & \cdots, \ \ 
\O ^{(1)} = - \frac{1}{2 p_0} \beta \left\{\left[\O , \E \right] + 
i\hbar \ddz \O \right\} + \cdots  
\eea 
It is to be noted that the transformation (\ref{FW1}) keeps the upper 
components of the beam optical wavefunction large compared to its lower 
components.  One can proceed with further FW-like transformations and stop at 
any desired stage.  Let us denote the $2$-component spinor comprising the upper 
components of the final $4$-component spinor obtained in the above process as 
$|\tilde{\psi}\rangle$.  

Up to now, all the observables, the field components, time etc., have been 
defined in the laboratory frame.  The covariant description of the spin of the 
Dirac particle has the simplest operator representation in the rest frame of the 
particle.  Thus, in accelerator physics the spin is defined in the rest frame 
of the particle.  So we make a further transformation which takes us from the 
beam optical form to the {\em accelerator optical form} 
\beq   
\left| \psi^{(A)} \right\rangle 
= \exp \left\{- \frac{i}{2 p_0} \left( 
\hat{\pi}_x \sigma_y - \hat{\pi}_y \sigma_x \right) \right\} \left| 
\tilde{\psi} \right\rangle\,. 
\eeq
Thus, up to the paraxial approximation the accelerator optical 
Hamiltonian~\cite{CJKP,JK3} is
\bea 
i \hbar \ddz \left| \psi^{(A)} \right\rangle
& = & \hH ^{(A)} \left| \psi^{(A)} \right\rangle\,, \nn \\ 
\hH ^{(A)} 
& \approx & \left(- p_0 - q A_z + \frac{1}{2 p_0} \pitwo \right) 
+ \frac{\g m}{p_0} \Vomeg \cdot \S \,, 
\eea  
where $\Vomeg = - \frac{1}{\g m} \left\{q \B + \eps 
\left( \B _\parallel + \g \B _\perp \right) \right\}$, $\g = E/mc^2$ and 
$\eps = 2m\mu_a/\hbar$.

\medskip

\section{An example of application: Magnetic quadrupole lens}
Let an ideal normal magnetic quadrupole of length $\ell$, characterized 
by the field $\B = (-Gy , -Gx , 0)$, be situated between the transverse planes 
at $z = \zi$ and $z = \zo = \zi + \ell$.  The associated vector potential
can be taken to be 
$\A = \left( 0 , 0 , \half G \left( x^2 - y^2 \right) \right)$  
with $G$ as constant inside the lens and zero outside. The accelerator optical 
Hamiltonian~\cite{CJKP} is 
\beq 
\hH (z) = \left\{ 
\ba{l}
\hH _F = - p_0 + \frac{1}{2 p_0} \ptwo \,, 
\quad {\rm for}\ \ z < \zi \ \ {\rm and}\ \ z > \zo \,,\\ 
\hH _L(z) = - p_0 + \frac{1}{2 p_0} \ptwo - \half q G \left( x^2 - y^2 \right) 
+ \frac{\eta p_0}{\ell} \left( y \sigma_x + x \sigma_y \right)\,, \\ 
\qquad \qquad {\rm for}\ \ \zi \leq z \leq \zo \,, \quad 
{\rm with}\ \ \eta = (q + \g \eps )G\ell\hbar/{2 p_0 ^2}\,.  
\ea \right. 
\eeq   
The subscripts $F$ and $L$ indicate the field-free and the lens regions 
respectively.

\medskip

Best way to compute the $z$-evolution operator $\hat{U}$ is {\em via} the 
interaction picture, used in the Lie algebraic formulation~\cite{Lie} of 
classical beam optics. Using the transfer operator thus derived~\cite{CJKP} 
we get the transfer maps for the averages of the transverse phase-space 
components: with the subscripts {\em in} and {\em out} standing for 
$(z_{\rm in})$ and $(z_{\rm out})$, respectively, 
\bea 
& & \left( 
\ba{c}
\langle x \rangle \\  \\ 
{\langle \hat{p}_x \rangle } /p_0 \\   \\
\langle y \rangle \\  \\ 
{\langle \hat{p}_y \rangle}/{p_0} \\   \\
\ea \right) _{\rm out}
\approx 
T_Q
\left( \left( 
\ba{c}            
\langle x \rangle \\   \\ 
{\langle \hat{p}_x \rangle}/{p_0} \\  \\
\langle y \rangle \\  \\ 
{\langle \hat{p}_y \rangle}/{p_0} \\   \\
\ea \right) 
+ \eta \left( 
\ba{c}
\left(\frac{\cosh\,\left(\sqrt{K}\,\ell \right) - 1}{K \ell} \right) 
\langle \sigma _y \rangle \\ 
- \left( \frac{\sinh\,\left(\sqrt{K}\,\ell \right)} {\sqrt{K}\,\ell}\right) 
\langle \sigma _y \rangle \\ 
- \left( \frac{\cos\,\left(\sqrt{K}\,\ell \right) - 1}{K \ell} \right) 
\langle \sigma _x \rangle \\ 
- \left( \frac{\sin\,\left(\sqrt{K}\,\ell \right)} {\sqrt{K}\,\ell}\right) 
\langle \sigma _x \rangle 
\ea \right) \right)_{\rm in}\,, \nn \\
& & \nn \\
& & 
T_Q = M_> M_Q M_<\,, \qquad \qquad 
M_{\stackrel{<}{>}} =
\left(
\ba{cccc}
1 & \Delta z_{\stackrel{<}{>}} & 0 & 0  \\
0 & 1 & 0 & 0 \\
0 & 0 & 1 & \Delta z_{\stackrel{<}{>}} \\
0 & 0 & 0 & 1 \\
\ea \right)\,, \\ 
 & & \nn \\ 
 & & M_Q =
\left( 
\ba{cccc}
\cosh (\sqrt{K} \ell) 
& \frac{1}{\sqrt{K}}\sinh (\sqrt{K} \ell) 
& 0 & 0 \\ 
\sqrt{K} \sinh (\sqrt{K} \ell) 
& \cosh (\sqrt{K} \ell)
& 0 & 0 \\ 
0 & 0 &
\cos (\sqrt{K} \ell) 
& \frac{1}{\sqrt{K}}\sin (\sqrt{K} \ell) \\   
0 & 0 &
- \sqrt{K} \sin (\sqrt{K} \ell) 
& \cos (\sqrt{K} \ell)
\ea \right)\,. \nn 
\eea 
Thus we have got a fully quantum mechanical derivation of the combined effect of 
the focusing action of the quadrupole lens (note the traditional transfer matrices) 
and the Stern-Gerlach force.  It may be noted that the quantum formalism of spinor 
beam optics supports, {\em in principle}, the idea of a Stern-Gerlach spin-splitter 
device to produce polarized beams.~\cite{Splitter}  The transfer map across the 
quadrupole lens for the spin components computed using the above accelerator 
optical Hamiltonian describes the well known Thomas-Bargmann-Michel-Telegdi spin 
evolution.~\cite{CJKP}   

\medskip

\section{Concluding remarks} 
In fine, we have seen how one can obtain the formalism of quantum beam optics 
for any particle, starting {\em ab initio} from the relevant basic quantum 
equation, at the single-particle level.  A two-step algorithm for this purpose has 
been suggested.  Using the general principle, the construction of a spinor 
theory of accelerator optics, starting from the Dirac equation and taking into 
account the anomalous magnetic moment, has been demonstrated.  As an example of 
application of the resulting formalism the normal magnetic quadrupole lens has 
been discussed.  In the classical limit the quantum formalism leads to the Lie 
algebraic formalism of charged-particle beam optics.  

\medskip

To get a formalism taking into account the multiparticle effects, particularly 
for the intense beams, it should be worthwhile to be guided by the 
quantum-{\em like} approaches to the particle beam transport: Thermal Wave 
Model~\cite{Fedele} and Stochastic Collective Dynamical Model~\cite{Cufaro}.  
Recently the quantum-{\em like} approach has been applied to construct a 
{\em Diffraction Model} for the beam halo.~\cite{Halo}  This model provides 
numerical estimates for the beam losses.  In this context, another useful 
approach could be to use the Wigner phase-space distribution functions.  
Heinemann and Barber~\cite{HB} have initiated the derivation of such a 
formalism for the Dirac particle beam physics starting from the original work 
of Derbenev and Kondratenko~\cite{DK} who used the FW technique to get their 
Hamiltonian for radiation calculations. 

\medskip

The present study is confined to systems with straight optic axis.  An 
extension to the curved optic axis systems should be done.  This would involve 
the subtlities of quantization in curvilinear coordinates.  Then there are the 
well known questions related to the position operator in the relativistic 
quantum theory.  Also, there are doubts about the exact form of the Stern-Gerlach  
force for a relativistic particle.~\cite{SG}  To address such questions from the 
point of view of experiments using particle beams the right platform would be 
the formalism of quantum beam optics. 

\medskip

\section*{Acknowledgments}
I am very grateful to Prof. R. Jagannathan, for all my training in the
exciting field of {\em quantum theory of charged-particle beam optics}.
I am thankful Prof. M. Pusterla for kind encouragement.  It is a pleasure to 
thank the Organizing Committee of QABP2K and Universit\`{a} di Salerno, for 
providing full financial support for my travel and stay to participate in QABP2K.

\end{document}